\documentclass[
reprint,
nofootinbib,
amsmath,amssymb,
aps,
prstper,
floatfix,
]{revtex4-2}

\usepackage{graphicx}
\usepackage{dcolumn}
\usepackage{bm}
\usepackage{hyperref}

\usepackage{tabularx}
\usepackage{makecell}
\usepackage{array}

\begin{document}


\title{From A Systematic Investigation of Faculty-Produced Think-Pair-Share Questions to\break
Frameworks for Characterizing and Developing Fluency-Inspiring Activities}

\author{Rica Sirbaugh French}
\email{rfrench@miracosta.edu}
\homepage{https://tiny.cc/rfrenchmcc}
\affiliation{Department of Physical Sciences; MiraCosta College; 1 Barnard Drive; Oceanside, CA 
18
 92056, USA}
\affiliation{Center for Astronomy Education, Department of Astronomy, Steward Observatory; University of Arizona; Tucson, AZ 85721, USA}


\author{Edward E. Prather}
\email{eprather@email.arizona.edu}
\affiliation{Center for Astronomy Education, Department of Astronomy, Steward Observatory; University of Arizona; Tucson, AZ 85721, USA}

\date{January 15, 2020}

\begin{abstract}
Our investigation of 353 faculty-produced multiple-choice Think-Pair-Share questions leads to key insights into faculty members' ideas about the discipline representations and intellectual tasks that could engage learners on key topics in physics and astronomy. The results of this work illustrate that, for many topics, there is a lack of variety in the representations featured, intellectual tasks posed, and levels of complexity fostered by the questions faculty develop. These efforts motivated and informed the development of two frameworks: (1) a \textit{curriculum characterization framework} that allows us to systematically code active learning strategies in terms of the discipline representations, intellectual tasks, and reasoning complexity that an activity offers the learner; and (2) a \textit{curriculum development framework} that guides the development of activities deliberately focused on increasing learners' discipline fluency. We analyze the faculty-produced Think-Pair-Share questions with our curriculum characterization framework, then apply our curriculum development framework to generate (1) \textbf{Fluency-Inspiring Questions}, a more pedagogically powerful extension of a well-established instructional strategy, and (2) \textbf{Student Representation Tasks}, a brand new type of instructional activity in astronomy that shifts the responsibility for generating appropriate representations onto the learners. We explicitly unpack and provide examples of Fluency-Inspiring Questions and Student Representation Tasks, detailing their usage of \textbf{Pedagogical Discipline Representations} coupled with novel question and activity formats.
\end{abstract}

\maketitle


\section{\label{sec:intro}Introduction}

In this article we describe how an investigation into faculty-produced curricular materials provides unique insights into the choices instructors make when designing their own active learning strategies. The results of this investigation also expand the theoretical underpinnings that inform our current curriculum development efforts. This investigation focuses on 353 multiple-choice questions authored by faculty (during professional development workshops) to target students' conceptual and reasoning difficulties associated with commonly taught topics in introductory astronomy and physics. Our goal was to characterize the information faculty choose to include and emphasize in their questions, and better understand how they structure intellectual tasks they believe will help foster rigorous discourse among students during Think-Pair-Share \cite{BP,PB2009} (or ``Peer Instruction'' as it is commonly referred to in the college-level introductory physics community \cite{Mazur1997}). After an initial analysis of the questions, it was clear that we needed a rigorous methodology that would allow us to meaningfully characterize the rich information contained within these questions. Informing our investigation with the theory of social semiotics (the study of communication and meaning-making potentials via signs and symbols that are highly contextualized within a community) has proven especially valuable. We will describe the professional development experiences during which these faculty-produced questions were created and provide insights into our findings on the discipline representations and cognitive tasks used in the questions.

Additionally, we introduce two exciting new frameworks: a \textit{curriculum characterization framework} and a \textit{curriculum development framework}. Arising from our systematic, socio-semiotic analysis of the faculty-produced Think-Pair-Share questions, our \textit{curriculum characterization framework} allows consistent, objective characterization of the information contained within any piece of curriculum, set of instructional materials, active learning strategy, etc. We outline its three-pronged design and offer examples of its application to the faculty-produced multiple-choice questions. Informed by and extending this work, our \textit{curriculum development framework} guides the creation of instructional strategies that employ novel combinations of discipline representations and intellectual tasks designed to help learners' develop their discipline fluency. We present \textbf{Fluency-Inspiring Questions} and \textbf{Student Representation Tasks} as two curricular objects generated from implementing our curriculum development framework. While Fluency-Inspiring Questions expand upon a well-established instructional strategy, Student Representation Tasks are unlike anything previously developed for instruction in astronomy as they ``flip the script,'' requiring the learners (rather than the instructors) to create the appropriate representations.

In the final sections of this paper we provide and explicitly unpack example Fluency-Inspiring Questions and Student Representation Tasks, highlighting how we as instructors can shift our thinking towards designing more pedagogically powerful materials and learning experiences that can foster learners' development of discipline fluency. In the next section we offer some background on our prior curriculum development efforts and theoretical perspectives.

\section{\label{sec:background}Background and Theoretical Underpinnings}

For the past two decades, the authors and their collaborators at the Center for Astronomy Education (CAE) have conducted research on the development and effectiveness of active learning instructional strategies and assessment materials primarily for use in general education introductory astronomy courses (commonly referred to as ``Astro 101'' \cite{French2019,RPBCG2010,PRB2009}). All of these efforts are informed by a multitude of different theoretical perspectives including, but not limited to, constructivism \cite{vonGlaserfeld1995,Vygotsky1978,Piaget1964,Bruner1960}, conceptual change theory \cite{DT2003,SAD1991,PSHG1982}, cognitive load theory \cite{CS1991}, ontological categories \cite{Chi2008,CR2002}, phenomenological primitives and knowledge in pieces \cite{PSO2003,diSessa1988}, activation of resources \cite{HESR2006}, facets of knowledge \cite{Minstrell1992}, and variation theory \cite{Lo2012,MP2006}. These theoretical perspectives significantly influence the design of \textit{all} our instructional materials to ensure they (1) are sensitive to the complexities of the educational contexts as well as students' ideas, prior knowledge, and intellectual abilities, (2) offer a variety of representations and scenarios that provide developmentally appropriate access to the topics, (3) foster critical and reflective thinking, (4) promote meaningful peer-to-peer discourse, and (5) engage students in a variety of cognitive tasks (e.g. draw, write, rank, sort, predict, calculate, etc.). Examples of materials informed by these theoretical perspectives include Lecture-Tutorials \cite{WP2019,WCPB2016,WPHBSC2016,PSQBJD2005}, Ranking Tasks \cite{HPGS2007}, Think-Pair-Share questions \cite{CPB2011,PB2009}, and concept inventories \cite{WWP2013,BJPS2012,WPD2011ptI,WPD2011ptII,BPBS2007}. See references \cite{WCP2018,EPW2015,SPWRB2012,WPD2012ptIII,WPD2012ptIV,WPD2012ptV,L2010,PRBS2009} for research into the development and assessment of these active learning strategies and assessment materials.

\subsection{\label{sec:PDRs}Pedagogical Discipline Representations}

Here, we outline a particular orientation of our work with regard to curriculum development and the use of representations. The hierarchical and scaffolded nature of the learning sequences fostered by our instructional strategies requires that we break down complex astrophysical concepts into smaller more manageable chunks in terms of both content and cognitive load. This ``chunking'' and sequencing allows novice learners to process and coordinate the discipline information in ways that effectively facilitate the development of coherent explanatory mental models. As we endeavor to bring more advanced topics and recent discoveries in astronomy and astrophysics into the classroom, we frequently find ourselves needing a new generation of representations, ones that emphasize information in ways not typically employed in the discipline. These new representations depict stylized physical scenarios and highlight discipline relationships that, while invaluable pedagogically, have little to no value to experts and professionals working in that field. Generally speaking, the higher the pedagogical value of a representation, the lower its value to discipline experts \cite{AL2017,Airey2015}. For this reason, these new representations are called \textbf{Pedagogical Discipline Representations} (PDRs) \cite{WCPB2016}.

Each Pedagogical Discipline Representation affords learners access to discipline information in ways that most textbook and expert representations simply cannot. More precisely, PDRs are \textit{representations} (ways of conveying discipline information) with specific, narrowly focused and well-understood disciplinary \textit{ affordances} (potentials for allowing access to pieces of disciplinary information) that promote \textit{unpacking} (disassembling a package of information and making the various pieces and connections explicit) while enabling critical and disciplinary \textit{discernment} (coming to recognize and understand what to focus on and interpreting it or making meaning using the appropriate context) \cite[and references therein]{U2019,AL2017,AEFL2014,FLAL2014,FAL2012}. For a more complete discussion on Pedagogical Discipline Representations and their development with respect to our work in astronomy see \citet{WCPB2016,WPHBSC2016} and \citet{HWPKC2018}. Our work on the development and testing of PDRs has significantly influenced the research described here to characterize the representations created by faculty members while developing their instructional strategies.

\subsection{\label{sec:semiotics}Social Semiotics}

The research and curriculum development described herein is strongly centered on better understanding how discipline representations and intellectual tasks are connected to the learning of a discipline. Viewing our work through the lens of social semiotics has been particularly helpful. As previously stated, social semiotics is the study of communication and meaning-making potentials via signs and symbols that are highly contextualized within a community, culture, etc. \cite{MODE2012,vanLeeuwen2005}. To facilitate making and conveying meaning, humans select and configure various \textit{modes of representation} -- channels for conveying information -- in complementary ways \cite{Jewitt2009}. These modes of representation (sometimes referred to as just ``modes'' or ``representations'' in this work) each have one or more \textit{affordances} -- potentials to allow access to components of information via an individual's perception of and interaction with the representation itself and a specific environment, context, etc. \cite{Gibson1979}.

Virtually every representation used in an instructional environment is limited by its own set of disciplinary affordances and pedagogical values \cite{U2019,AL2017,AEFL2014,FLAL2014,FAL2012}. Thus, elevating learners' knowledge of, and abilities in, the key ideas of a topic or discipline necessarily requires combining multiple representations coupled to multiple intellectual tasks in ways that facilitate unpacking and discerning \cite{AL2017}. Indeed, \citet{FLAL2014} and \citet{Linder2013} even suggest that the power and success of many research-validated active learning materials and methods may lie within a theoretical framing in which the method itself naturally facilitates the unpacking and disambiguation of the disciplinary affordances of a set of representations. Thus, studying instructional strategies by recognizing which combinations of representations and intellectual tasks are employed can provide insight into the pedagogical beliefs of the authors -- in our case, faculty.

Next, we describe our efforts to characterize the information contained within the hundreds of multiple-choice Think-Pair-Share questions produced by faculty participating in professional development workshops.

\section{\label{sec:TPSQs}Investigating faculty-produced {Think-Pair-Share} questions}

Our data is comprised of 353 multiple-choice Think-Pair-Share questions produced by faculty during professional development workshops that included a session designed to help instructors better understand how to create effective multiple choice questions and employ best practices for implementing Think-Pair-Share \cite{BP,PB2009}. Note that in these workshops, faculty were given experiences with evaluating and designing multiple-choice questions featuring an array of formats, levels of intellectual difficulty, and abilities to promote discourse amongst learners. 

The majority of the questions (293) are from 41 CAE Teaching Excellence Workshops \cite{PB2009,CAEworkshops} held from 2005-2015, targeting current and prospective instructors of college-level general education introductory astronomy, Earth, and space science, regardless of experience or career stage. These questions span the broad topical areas of the Earth-Sun-Moon system, Renaissance astronomy, solar system, light and atoms, stars, exoplanets and life in the universe, and galaxies and cosmology. The remaining 60 questions are from four meetings of the Workshop for New Faculty in Physics and Astronomy \cite{NFW} held at the American Center for Physics from 2015-2017. These professional development experiences support primarily physics and astronomy instructors who are in the first few years of their initial tenure-track appointments. Questions from these workshops fall under the broad topical areas of work and kinetic energy, inelastic collisions, rotational motion, heat and temperature, Gauss's law for electric fields, Faraday's and Lenz's laws, simple harmonic motion, and the Bohr model of the atom.

Through the careful investigation of these questions, we hoped to gain valuable insights into the choices faculty make when developing instructional materials they believe will help students learn their discipline. What aspects of a particular topic do faculty think are important? How do faculty choose to represent and emphasize certain pieces of information? What kinds of cognitive exercises do faculty think learners should experience? Do the questions faculty develop address the intended learning outcome(s)? 

From a cursory analysis of the data, we suspected there was a lack of variety in the representations, intellectual tasks, and levels of difficulty. But without an explicit framework to help us rigorously document the information contained in the questions, we could not be certain of the actual distribution of these question features. It was clear that we needed a more objective, insightful, and systematic way to characterize the abundant information contained within our data. Further, we believed the systematic use of an objective coding framework informed by social semiotics would help us better understand which representations and intellectual tasks are over- or under-utilized for a particular topic, and could inform directions for future curriculum development. The work of \citet{Linder2013} and \citet{AL2009} forms the basis for our initial efforts to develop a framework for systematically coding the modes of representation and intellectual tasks used, as well as the levels of discourse promoted in the faculty-produced multiple-choice questions. In the next section, we discuss the development of our \textit{curriculum characterization framework}.

\section{\label{sec:framework-char}Curriculum Characterization Framework}

Our \textit{curriculum characterization framework} is designed to meaningfully code the information contained in the multiple-choice questions using a three-pronged approach grounded in the answers to the following:
\begin{itemize}
\item{What types and how many different ways of conveying information are used?}
\item{What types of and how many different cognitive exercises must the learner engage in?}
\item{How robust will the discourse be among learners attempting to explain and defend the reasoning behind their answers?}
\end{itemize}

To address the first question, we developed a coding schema -- heavily influenced by \citet{Linder2013} and \citet{AL2009} -- that allows us to classify how information is represented. Identifying and systematically categorizing the different ways that faculty choose to convey information offers a key insight into understanding how faculty perceive the various representations' affordances and intended ways of making meaning, which are core components of socio-semiotic theory. To address the second question, we extended the socio-semiotic perspective on representations by creating a second coding schema for cataloging the different types of intellectual tasks one must engage in when working through an activity. For the third question, we refined an existing rubric (previously created by members of CAE) known as the Question Complexity Rubric (QCR) \cite{CPB2011}. This modified QCR is used to rank the complexity involved in unpacking, explaining, and justifying one's reasoning.

We arrived at final versions of the Question Complexity Rubric and coding schemata for the representations and intellectual tasks via an iterative process. We assembled a small, but reasonably diverse set of the faculty-produced Think-Pair-Share questions (referred to as the calibration questions) that exemplified a range of representations, tasks, and QCR codes. Using this set of calibration questions, we trained a team of collaborators in the application of the coding schemata and Question Complexity Rubric. Through an iterative process of coding the calibration questions and reflecting on our results, we made important revisions that led to the final versions of the coding schemata and QCR presented here. Subsequent efforts by the team to code additional questions confirms that our coding methodology leads to reliable and valid characterization of the data.

The three-pronged approach above, the coding schemata for the modes of representation and intellectual tasks, and the modified Question Complexity Rubric together comprise our \textit{curriculum characterization framework}. We describe the coding schemata and QCR in more detail in the following subsections.

\subsection{\label{sec:modes}Modes of representation}

We define a \textit{mode of representation} as a way of conveying information. The numbered modes of representation in Table~\ref{tab:modes} are general types of information delivery widely used in teaching college-level science. Some modes have lettered subtypes that serve as common examples and are included to assist in characterizing the general mode of representation. While Think-Pair-Share questions typically do not make use of modes 7--10 they are still included since our framework is easily generalized to other types of curricular materials and instructional methods.

\begin{table}[htbp]
\caption{\label{tab:modes}Modes of representation recognized in curricular materials.}
\begin{ruledtabular}
\begin{tabularx}{\linewidth}{dX}
\multicolumn{2}{c}{Modes of Representation}\\
\colrule
1. & words\\
   & a. written\\
   & b. spoken\footnote{\label{foot:spokenmode}Real-time only; does not include recordings; see mode 9.}\\
2. & pictures \& diagrams\\
   & a. photographs\\
   & b. static images\\
   & c. figures\\
   & d. sketches\\
3. & graphs \& charts\\
4. & tables\\
5. & mathematical formalism\footnote{Equations and other mathematical expressions, e.g. $\lambda_{\textrm{max}} = 6000 \mbox{\AA}$. Does not include ranked answer choices.}\\
6. & numbers\footnote{Used anywhere except items explicitly covered by mode 5.}\\
7. & animations\footnote{Moving pictures or diagrams with no user interaction. May include pause/resume/restart controls but disallows changing any variables or parameters.}\\
8. & simulations\footnote{Animated tool with user interaction mechanisms. User has the ability to change and/or control one or more variables or parameters.}\\
9. & recordings of reality\\
   & a. audio\\
   & b. video\\
10. & gestures\footnotemark[1]\\
   & a. facial expressions\\
   & b. body movements\\
\end{tabularx}
\end{ruledtabular}
\end{table}

Each question is coded by identifying the various modes of representation used and listing their corresponding numbers from Table~\ref{tab:modes} (see Table~\ref{tab:coding}). The order of the numbers is irrelevant but we often list them in the order the representations are encountered when working through the question.

\subsection{\label{sec:tasks}Intellectual tasks}

We define an \textit{intellectual task} as a specific cognitive exercise that one engages in to arrive at the answer to a question. Our list of task codes appears in Table~\ref{tab:tasks}.

\begin{table}[b]
\caption{\label{tab:tasks}Intellectual tasks recognized in curricular materials.}
\begin{ruledtabular}
\begin{tabularx}{110pt}{d >{\raggedright\arraybackslash}X d@{} >{\raggedright\arraybackslash}X}
\multicolumn{4}{c}{Intellectual Tasks}\\
\colrule
 1. & visualize   & 9.  & rank\\
 2. & draw/sketch & 10. & sort\\
 3. & model       & 11. & match\\
 4. & compare     & 12. & quantitative reasoning\\
 5. & identify    & 13. & calculate\\
 6. & predict     & 14. & apply/analyze\\
 7. & extrapolate & 15. & write\\
 8. & count       & & \\
\end{tabularx}
\end{ruledtabular}
\end{table}

Some intellectual tasks are ubiquitous for virtually all questions, such as ``recall'' or ``interpret.'' Characterizing a question using such ubiquitous tasks does not contribute meaningfully to our understanding of the essential reasoning the question evokes, nor does it help distinguish differences among questions. Such tasks, therefore, are not included in our work. Think-Pair-Share questions do not make use of task 15 in Table~\ref{tab:tasks} but, just as with modes 7--10 in Table~\ref{tab:modes}, it is included because our framework is easily generalized beyond the current application.

Unpacking the particulars of these intellectual tasks helps clarify how one categorizes the Think-Pair-Share questions in terms of the cognitive exercises they might promote. To \textit{visualize}, one makes a mental image using information in the provided representations, e.g. from a description, graph, table, etc. The task \textit{draw/sketch} is the act of creating a pictorial or diagrammatic representation or simply adding detail to a pre-existing one. Here, \textit{model}ing, means making a physical representation by gesturing and/or using props. To \textit{compare} is to make explicit use of similarities and/or differences by considering items, situations, etc. in relation to each other, e.g. which one is hottest. The task \textit{identify} means one has distinguished or recognized a single item, case, etc. as fitting one or more characteristics, criteria, etc., e.g. which of the labeled locations in the Hertzsprung-Russell diagram is a white dwarf. To \textit{predict} means to forecast one or more future events from the given information while \textit{extrapolate} is to estimate a value outside a given range by assuming known a trend extends accordingly. Determining how many items fit certain criteria is a \textit{count}ing task while \textit{rank} means arranging items in a certain order based on criteria such as hottest to coldest, greatest to least, etc. Classifying or separating items into categories or bins is a \textit{sort}ing task. To \textit{match} is to assign items to corresponding other items. \textit{Quantitative reasoning} means doing numerical reasoning using analytical and mathematical thinking (e.g. proportional reasoning) while \textit{calculate} means one determines a precise numerical value. The \textit{apply/analyze} task involves developing a line of reasoning and drawing a conclusion or reaching a decision using discipline-specific relationships, rules, laws, etc. This task is used when other listed tasks (from Table~\ref{tab:tasks}) \textit{cannot completely characterize all of the cognitive exercises one could engage in to reason through the question and arrive at an answer}. And finally, \textit{write} means creating a coherent narrative and is typically prompted by something like ``describe,'' ``explain,'' etc.

Most questions naturally require multiple intellectual tasks so it is necessary to differentiate between the main \textit{overarching} task and any supplementary \textit{supporting} tasks that might be needed to arrive at an answer. The wording of some questions automatically reveals the overarching task. For example, ``How many of the following...'' signals \textit{count}, while a question that asks for objects to be arranged in a particular order implies that \textit{rank} is the overarching task. Supporting tasks are those that, while not the main thrust of the question, are still likely to occur when doing the reasoning necessary to arrive at an answer. Tasks such as \textit{quantitative reasoning}, \textit{compare}, and/or \textit{visualize}, for instance, might necessarily precede the overarching task \textit{rank} and therefore must be included as supporting tasks. Any task listed in Table~\ref{tab:tasks} could serve as either a supporting task or the overarching task.

The list of supporting tasks for any given question should \textit{include all tasks that one could reasonably expect a learner in the target population \underline{might} engage in to arrive at an answer, \underline{whether they actually do so or not}}. For example, a question coded with the supporting tasks \textit{visualize}, \textit{model}, and \textit{sketch} does not necessarily mean that all three tasks are required or done by a learner when trying to answer the question. Rather, where one learner might find it sufficient to only \textit{visualize}, a different learner may need to \textit{model} and/or \textit{sketch} instead of, or in addition to, \textit{visualize}. Still, all three tasks must be included when coding that particular question.

A question is coded by first identifying the overarching intellectual task required to answer the question, identifying all reasonably possible supporting tasks, and then listing their corresponding numbers from Table~\ref{tab:tasks} (see Table~\ref{tab:coding}). We list the overarching task first with the supporting tasks following in parentheses. While the order of the supporting tasks is not relevant, our lists often correlate with the order of tasks that a learner might engage in when formulating an answer to the question.

\subsection{\label{sec:QCR}Question Complexity Rubric ({QCR})}

\begin{figure}[b]
\includegraphics[width=\linewidth]{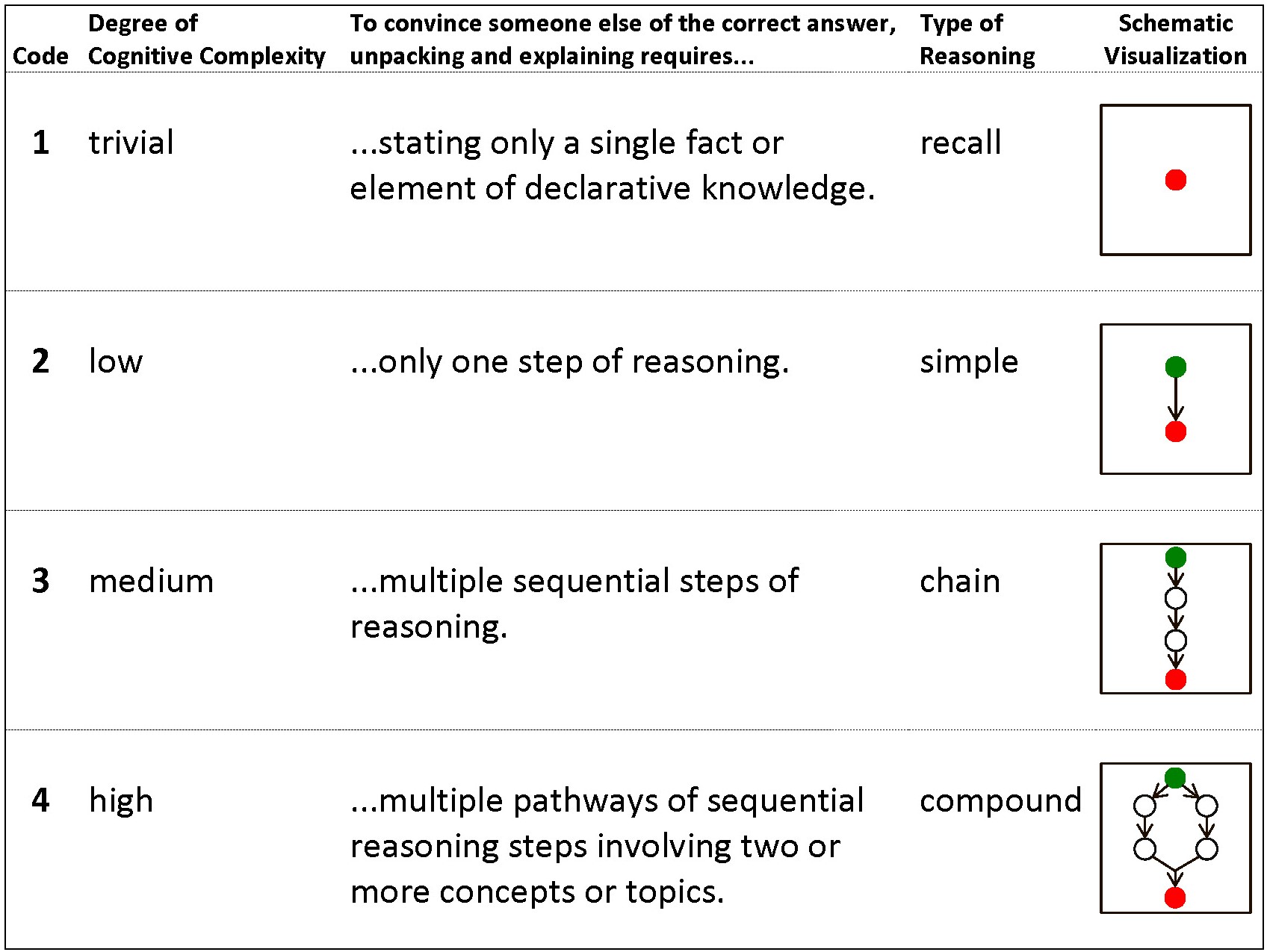}
\caption{\label{fig:QCR}The Question Complexity Rubric (QCR) used to code the level of cognitive complexity required to unpack and defend the solution to a question, problem, etc.}
\end{figure}

Building upon the initial work of \citet{CPB2011}, a question's QCR code (Fig.~\ref{fig:QCR}) ranks the question's degree of conceptual and cognitive complexity -- a ranking which also characterizes the richness of the conversation intended to be evoked between learners attempting to explain and defend the reasoning behind their answers. In this way, the QCR code represents the level of intellectual engagement required to convince someone else of the correct answer.

Thus, to determine a question's QCR code we consider what it takes to unpack and make explicit the affordances, pieces of knowledge, and reasoning necessary to develop and articulate a coherent narrative that should convince another learner of the correct answer. We then consider whether there are multiple concepts that must be integrated together while reasoning through the question. Each question is coded by assigning a number, 1 through 4 (the QCR code), from Fig.~\ref{fig:QCR} (see Table~\ref{tab:coding}). For a question coded as QCR~= 1, the learner needs only to state a single fact or element of declarative knowledge, whereas a question coded as QCR~= 4 requires the learner to defend their answer using multiple pathways of sequential reasoning steps and integrate two or more concepts or topics.

\section{\label{sec:data}Data characterization}

Using our \textit{curriculum characterization framework}, we systematically coded the information contained in each of the 353 multiple-choice Think-Pair-Share questions written by faculty participating in professional development workshops. While it is not within the scope of this paper to unpack the breadth and details of all of the data (a separate paper for this is in preparation), we show two questions and offer a few key findings.

\begin{figure*}[htbp]
\includegraphics[width=\linewidth]{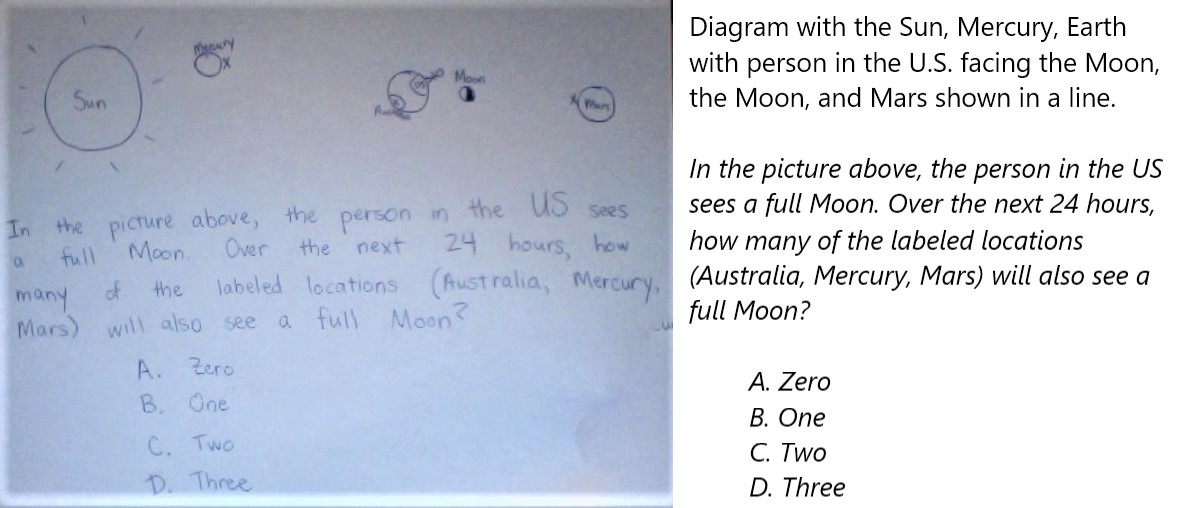}
\caption{\label{fig:sampleQ01}Example A from the faculty-produced Think-Pair-Share questions. The original is reproduced on the left with a ``transcript'' on the right. Italicized portions indicate text of question prompt and answer choices.}
\end{figure*}

\begin{figure*}[htbp]
\includegraphics[width=\linewidth]{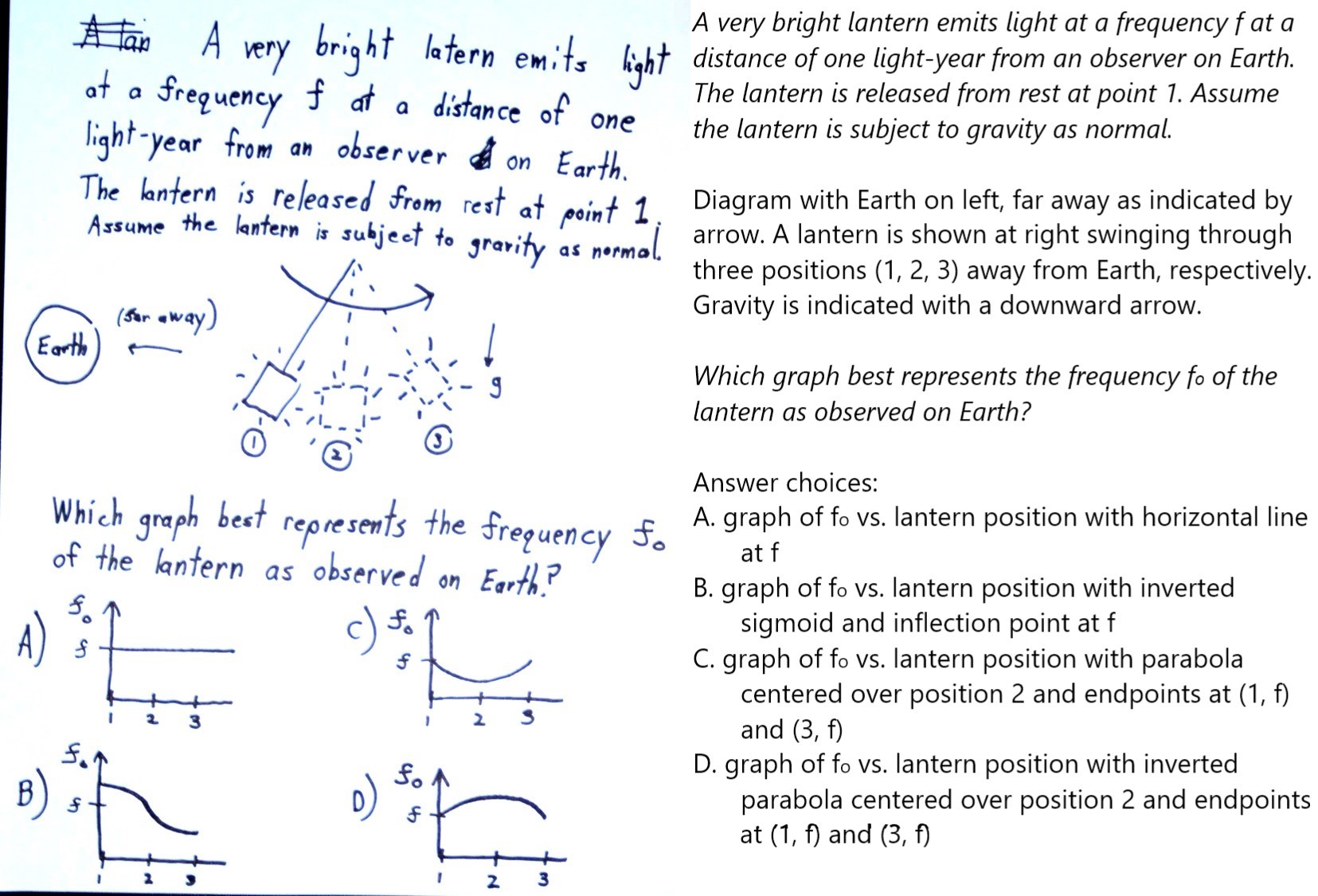}
\caption{\label{fig:sampleQ02}Example B from the faculty-produced Think-Pair-Share questions. The original is reproduced on the left with a ``transcript'' on the right. Italicized portions indicate text of question prompt.}
\end{figure*}

Figures~\ref{fig:sampleQ01} and \ref{fig:sampleQ02} 
each present a faculty-produced question from our data set. In both figures, the left side preserves the representations \textit{exactly} as presented by the faculty who created them while the right side shows a ``transcript'' of the question with italics indicating exact text from the question and/or answer choices. Table~\ref{tab:coding} contains our codes for these questions' representations, intellectual tasks, and QCR rankings. We chose these two questions because they highlight how multiple-choice questions can effectively utilize multiple representations coupled with several intellectual tasks to require complex multi-step reasoning -- a combination that was not common in the questions in our data set.

\begin{table*}[htbp]
\caption{\label{tab:coding}Results of coding the questions in Figs.~\ref{fig:sampleQ01} and \ref{fig:sampleQ02} using our curriculum characterization framework from \S\ref{sec:framework-char}.}
\begin{ruledtabular}
\begin{tabular}{ccccc}
\makecell{Example\\Question} & Broad Topic Area & \makecell{Modes of\\Representation\footnote{Note that while we indicate only the numbers of the modes and tasks when coding our data, we include the corresponding names of the modes and tasks here in the table footnotes, to aid the reader.}} & \makecell{Intellectual Tasks:\\overarching (supporting)\footnotemark[1]} & \makecell{QCR\\Code}\\
\colrule
A from Fig.~\ref{fig:sampleQ01} & Earth-Sun-Moon system & 2, 1, 6\footnote{pictures/diagrams, words, numbers} & 8 (14, 6, 1, 2, 3)\footnote{count (apply/analyze, predict, visualize, draw/sketch, model)} & 4\\
B from Fig.~\ref{fig:sampleQ02} & light and atoms       & 1, 2, 6, 3\footnote{words, pictures/diagrams, numbers, graphs/charts} & 11 (1, 3, 4, 12)\footnote{match (visualize, model, compare, quantitative reasoning)} & 4\\
\end{tabular}
\end{ruledtabular}
\end{table*}

Having worked with thousands of physics and astronomy faculty over the years in professional development settings, we find that many believe it is extremely difficult to use multiple-choice questions to engage learners in higher-order thinking and reasoning (i.e. the upper levels of Bloom's taxonomy \cite{AKB2001,BEFHK1956}). Somewhat unsurprisingly then, after applying our curriculum characterization framework to all of the questions in the data, we find few with a QCR code of 4. Out of the entire set of 353 faculty-produced questions spanning 15 broad topical areas, 1.7\% are QCR~= 1, 25.5\% are QCR~= 2, 56.9\% are QCR~= 3, and only 15.9\% are QCR~= 4. Pedagogically speaking, this result is problematic. Helping learners develop a more robust understanding of a topic involves scaffolding their learning, starting with novice-level situations and working up to problems that feature a wide variety of representations coupled with complex reasoning tasks that promote expert-like thinking. If multiple-choice questions are a primary source of engagement with, and assessment of, a particular topic, there must be a diverse assortment of questions that span all QCR levels for that topic. Our data suggests that faculty may be unlikely to produce a significant number of QCR~= 4 questions spanning the vast array of topics they are likely to address over a term of physics or astronomy instruction. It is our experience that faculty often utilize far too many low-level questions when incorporating Think-Pair-Share into the classroom. Through our efforts described here we are explicitly working to expand our communities' capacities to identify and develop QCR~= 4 questions and activities.     

We also found that there are some topics from this data whose questions are notably lacking in the diversity of representations and/or tasks. For example, the 67 questions in the ``stars'' topical area show a considerable over-reliance on the Hertzsprung-Russell diagram. This diagram, and the words that frame the question and context of the problem, are frequently the only representations used. It is also worth noting that the Hertzsprung-Russell diagram, while virtually indispensable, is arguably one of the most rationalized\footnote{\label{foot:rationalize}Representations commonly used to teach a particular concept (like those found in textbooks) frequently harbor key disciplinary aspects that are not immediately discernible. These ``rationalized'' representations contain information that has been compartmentalized via extensive discussion and reconciliation by discipline experts, often over long periods of time \cite{FLAL2014}.} and information-rich representations in all of astronomy, so instructors must be wary of trivializing its significant depth \cite{AE2019,ERR2017,Brogt2009}. Additionally, these same ``stars'' questions show an overwhelming preference for the overarching intellectual task \textit{compare}, with little variation in supplementary tasks. Similarly, most of the 75 questions on ``galaxies and cosmology'' use words as the lone representation and emphasize \textit{identification} as the overarching intellectual task, sometimes with no supporting tasks.

The insights gained from developing and applying our curriculum characterization framework to these faculty-produced Think-Pair-Share questions drove us to (1) systematically identify gaps in the diversity of modes and tasks, (2) generate questions to fill those gaps, (3) generate more complex questions that combine multiple representations and tasks in ways not seen in the data, and (4) create more pedagogically interesting questions that guide the unpacking of compound and/or complex ideas for the learners rather than leave them to start from a ``blank slate.'' Our \textit{curriculum development framework}, described next, was informed by this work.

\section{\label{sec:framework-dev}Curriculum Development Framework}

We take ``fluency'' in a discipline idea to mean the ability to easily unpack and transition through multiple modes of representation (including \textit{generating} them when necessary), discern their disciplinary affordances, and engage in various cognitive exercises to develop, apply, and articulate meaning in the appropriate disciplinary context(s). Thus, successful pedagogies must (1) acknowledge the disciplinary affordances of multiple complementary representations, (2) integrate them with multiple different intellectual tasks, (3) organize the information appropriately, and (4) offer plentiful opportunities for learners to practice unpacking, discerning, making meaning, and articulating reasoning while fostering reflection and self-assessment. When this host of processes becomes unproblematic and nearly second-nature or automatic, one is said to be fluent \cite{AL2017}.

Experts in a discipline routinely engage in ``disciplinary discourse'' \cite{Airey2011}, moving with practiced skill among a variety of representations -- selecting, interpreting, explaining, reconciling, and generating them -- within a context that is updated as discipline knowledge and understanding progresses. Facilitating the development of similar ``representational competence'' \cite{LAMW2014} in novice learners is challenging, even more so since the discipline content is often perceived as independent of the representations used \cite{KF2017}. Novices lack the ability to recognize important information and relationships in a variety of modes and filter it through the appropriate context(s). That is, they cannot yet critically discern the disciplinary affordances of multiple representations and coordinate them to make sense of disciplinary knowledge \cite{ELAR2014b}.

\citet{KF2017} argue that developing fluency in disciplinary content cannot be separated from the representations used to teach that content. We concur with their finding that requiring students to engage in multimodal\footnote{\label{foot:multimodal}In social semiotics, the term ``multimodality'' is often applied to the use of multiple semiotic resources, e.g. modes of representation, and their affordances working together to create an instance of communication \cite{MODE2012,Jewitt2009,KvL2001}.} learning can improve learners' performances. \citet{AL2009} point out that learners who have not been exposed to a wide variety of representations for a topic are unlikely to become fluent in that topic. Therefore, instructors who employ a limited set of representations cannot expect to move their students to discipline fluency since the learners' opportunities to unpack and discern are restricted by the lack of variety. Additionally, students tend to call upon the representation(s) most frequently used rather than the one(s) best suited to the task at hand \cite{FAL2012}.

Since different representations offer different disciplinary affordances \cite{FLAL2014,McDermott1990}, no single representation by itself is likely to capture all aspects of the physical situation it models, regardless of rationalization.\footnotemark[1] Multiple representations, however, may work together to create a ``collective disciplinary affordance'' \cite{Linder2013}, providing a more holistic model of the physical situation and more potential points of access for the development of disciplinary knowledge and fluency. Thus, it is pedagogically more powerful to use multiple representations than to rely on a single one to do the work of many \cite{FLAL2014}. This is a guiding principle behind the new kind of multiple-choice questions we call \textbf{Fluency-Inspiring Questions} (see \S\ref{sec:FIQs}).

However, just because a student answers one or more questions correctly does not necessarily imply a deep disciplinary understanding indicative of fluency. Learners must also be able to unpack those resources \cite{FLAL2014} and come to notice, appreciate, and effectively coordinate the disciplinary affordances of those modes lest they fall victim to \textit{discourse imitation} -- the ability to utilize representations appropriately without having the associated discipline-specific understandings \cite{AL2017,AL2009}. This suggests the need for learners to also be able to \textit{generate} appropriate representations when needed, and not simply know how to utilize pre-existing ones. Such is the motivation behind the novel activities we call \textbf{Student Representation Tasks} (see \S\ref{sec:SRTs}).

Informed by these theoretical perspectives on discipline fluency and our results from applying our curriculum characterization framework, we created our \textit{curriculum development framework}. This framework is structured to aid the generation of learning opportunities that explicitly promote fluency and requires developers to: 
\renewcommand{\labelenumi}{(\arabic{enumi})}
\begin{enumerate}
\item{unpack a discipline topic in terms of the learning outcomes for the intended audience, e.g. discern what is required in order for one to demonstrate fluency;}
\item{examine the canonical discipline representations used for the topic to determine whether they have the appropriate affordances or whether they mislead or unintentionally create opaque learning environments that result in learners developing incomplete, incorrect, and/or incoherent ideas and explanations;}
\item{determine whether there are different modes of representation that could or should be used, and whether the creation of new Pedagogical Discipline Representations is necessary;}
\item{design complex scenarios including intellectual tasks that require the learner to use a combination of essential features from various modes of representation in order to develop robust explanations; and finally,}
\item{structure the activity such that it challenges learners at a high enough cognitive level to ultimately aid in developing and demonstrating fluency in the topic, i.e. QCR~= 4.}
\end{enumerate}

We next unpack \textbf{Fluency-Inspiring Questions} and \textbf{Student Representation Tasks}, two innovative instructional strategies generated from the application of our curriculum development framework.

\section{\label{sec:FIQs}Fluency-Inspiring Questions}

The multiple-choice questions we generate using our \textit{curriculum development framework} are \textbf{Fluency-Inspiring Questions} (FIQs) in that they require the learner to extract information from one representation and map it onto one or more others while engaging in multiple cognitive tasks. It is important to note that Fluency-Inspiring Questions are for use in a post-instruction context which, in our case, means post-lecture and usually after implementing other collaborative active learning strategies (such as Lecture-Tutorials). The following examples illustrate how the structures and designs of FIQs purposefully facilitate discernment, unpacking, and mapping of information from one representation to another, leading the learner to work through a series of complementary tasks and complex discipline-specific scenarios, resulting in a QCR code of 4. In particular, these examples employ unusual forms of fill-in-the-blank and matching, two unexpectedly powerful question formats that can help foster robust intellectual engagement that leads to more expert-like disciplinary discourse among learners.

\subsection{\label{sec:FIQ:heatengine}Heat engine}

\begin{figure}[htbp]
\includegraphics[width=\linewidth]{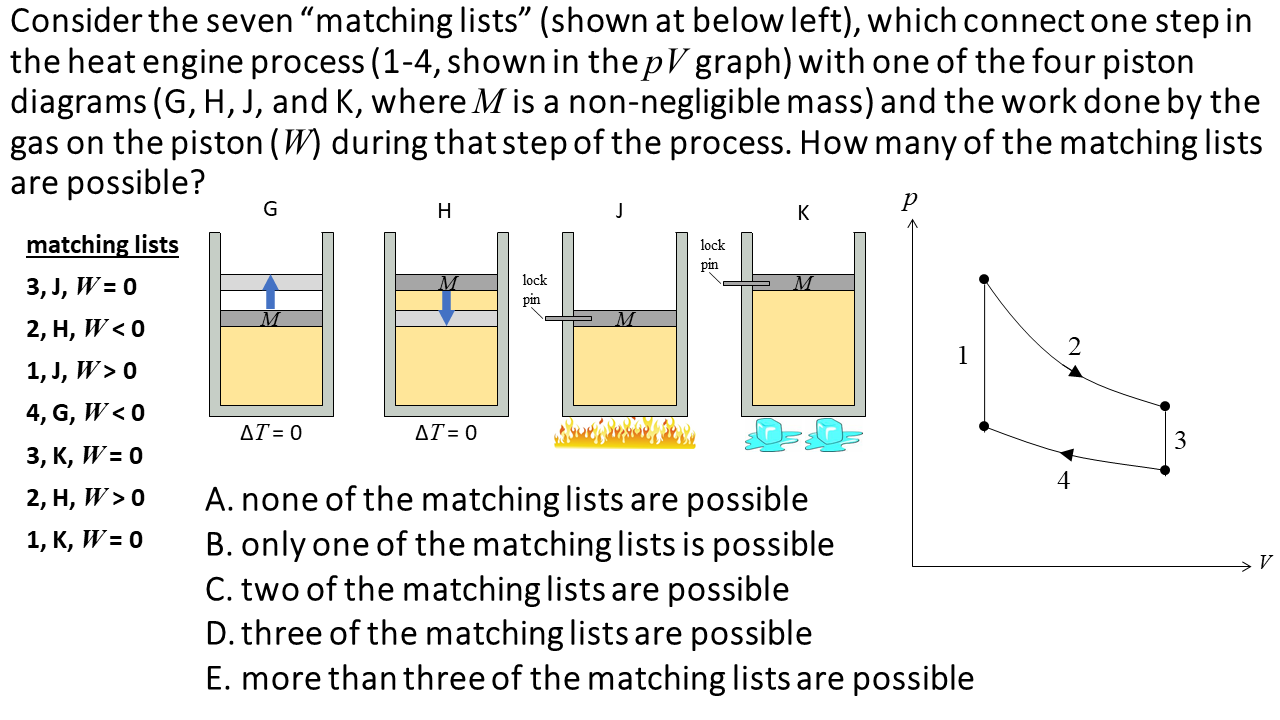}
\caption{\label{fig:FIQ:heatengine}A thermodynamics FIQ for a heat engine process.}
\end{figure}

When teaching thermodynamics, instructors have an expectation that learners will be able to correctly connect the ideas of work and energy transfer to the paths in a $pV$ graph and to sequences of real, physical events, such as those depicted in commonly used piston diagrams. Correctly connecting these ideas is a hallmark of demonstrating fluency with a heat engine process. What we commonly find, though, are questions that deal with one part of the graphical path and a partial sequence of the piston diagram process, or questions asking students to calculate the work for the displacement of the piston under a partial set of conditions. These questions still address only pieces of the larger, interconnected set of concepts. If we are not careful, creative, and deliberate in our question design, we may inadvertently inspire only discourse imitation \cite{AL2017,AL2009}. What we really need is a question that requires the learner to evaluate all aspects of a full thermodynamic cycle, connecting the theoretical processes, physical manifestations, graphical representations, and mathematical consequences \textit{simultaneously}. The Fluency-Inspiring Question in Fig.~\ref{fig:FIQ:heatengine} is designed to accomplish this.

Aside from the novel ``matching list,'' the representations used are all fairly common, with similar versions found in most introductory physics and thermodynamics textbooks and curricular materials. There are no Pedagogical Discipline Representations in this example and none are needed since the canonical representations afford access to the necessary pieces of disciplinary information. The use of a ``matching list'' with the overarching task \textit{count} sets up a curiously powerful combination of representations and tasks that orchestrates the learner's cognitive efforts by requiring one to map back and forth among a variety of pieces of discipline information while simultaneously calling upon multiple theoretical and mathematical principles. This results in a unique fluency-inspiring opportunity.

Thus, we have explicitly addressed the five requirements of our curriculum development framework (\S\ref{sec:framework-dev}). And while the individual pieces of this question are widely taught in introductory thermodynamics lessons, their combination in this format proves intellectually challenging -- sometimes even for faculty working through the problem during professional development workshops on the design of Think-Pair-Share questions.

\subsection{\label{sec:FIQ:bohr}The Bohr atom and {EM} radiation}

Instruction on the electron transitions inside an atom typically involves traditional energy level diagrams and assumes that students have (1) a functional understanding of the concepts of emission and absorption and (2) mastered the relationships between wavelength, frequency, and energy of a photon. Successful integration of these ideas indicates fluency with the interaction between photons and atoms. The properties of light and (sometimes) the processes of emission and absorption are often first dealt with in a more compartmentalized, piecewise fashion, and faculty often assume that students will automatically transfer and coherently reason about these ideas when encountering representations of energy levels in an atom. From our decades of classroom experience teaching these topics, we find that this is very often not the case. For example, for the cases of absorption and emission, students frequently confuse the direction of the arrow used to indicate the transition of the electron. Students also struggle to correctly connect the magnitude of the electron transition to the corresponding energy or wavelength of the associated photon, incorrectly reasoning that length of the electron transition arrow is directly proportional to the photon's wavelength, or that transitions involving a greater number of energy levels always result in (or from) a photon of greater energy.

\begin{figure}[b]
\includegraphics[width=\linewidth]{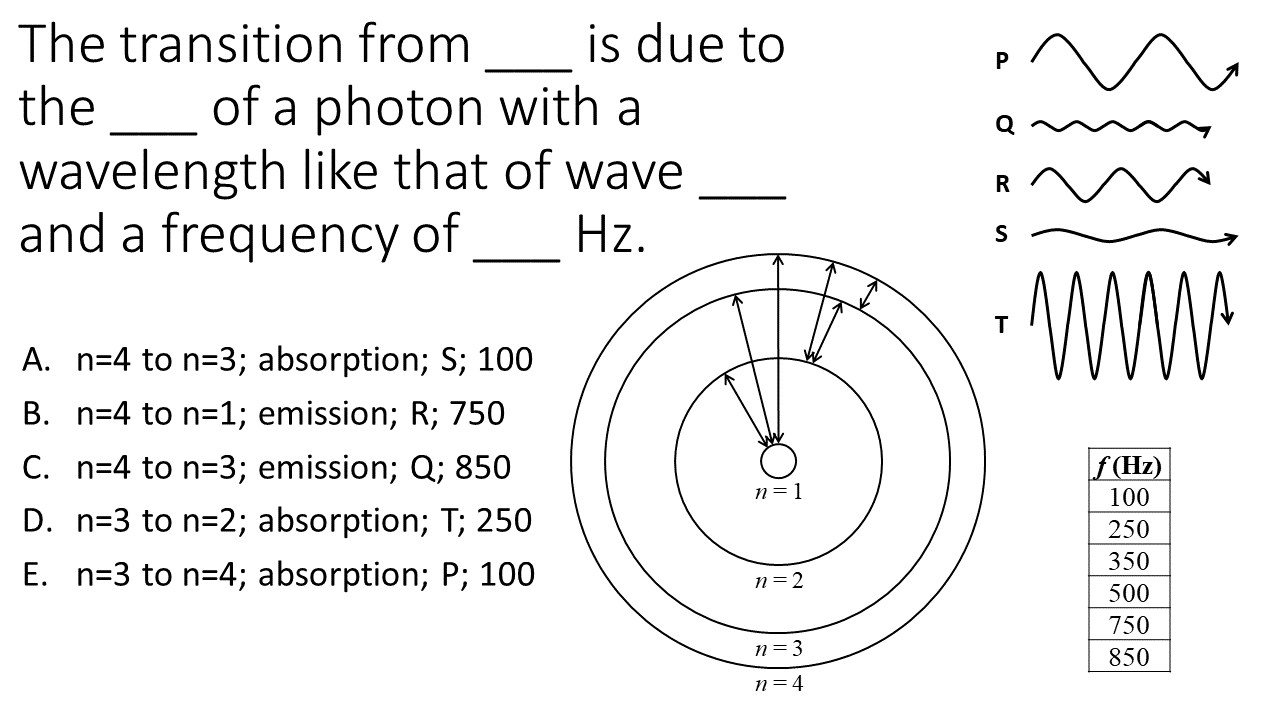}
\caption{\label{fig:FIQ:bohr}FIQ integrating electromagnetic radiation with the Bohr model of the atom.}
\end{figure}

The Fluency-Inspiring Question in Fig.~\ref{fig:FIQ:bohr} 
uses a fill-in-the-blank format to guide the students' thought processes. In this question we see aspects of traditional representations that have been altered to serve as Pedagogical Discipline Representations. For example, the Bohr atom PDR in Fig.~\ref{fig:FIQ:bohr} combines information about energy levels and all possible bound -- ``arrowed'' -- electron transitions with the productive orbital features of the Bohr atom in a somewhat unconventional way that affords learners access to the phenomena of emission and absorption and the corresponding movement of electrons between energy levels. Similarly, ``wiggly'' arrows are sometimes shown when textbooks introduce photons and electromagnetic waves. But combining multiple wiggly arrows of varying wavelength, frequency, and amplitude into a single diagram used to represent different physical parameters of photons in this way is uncommon in traditional astronomy and physics instruction. Yet it is invaluable when the differentiation of these photon properties is a goal of instruction on the topic. Thus, these wiggly arrows become a PDR. And while the frequencies in the table are unrealistic for an actual atom, they facilitate paying attention to the \textit{relationships} among the frequencies that result from the physical situation. This helps novice learners to focus on the appropriate relationships rather than determining precise mathematical values. Thus, this frequency table is also a PDR.

The format of this Fluency-Inspiring Question requires simultaneously mapping back and forth between the arrows on the Bohr atom, the properties of the wiggly photons, ideas about emission versus absorption, and photon frequencies and energies, thus elevating this question to a QCR~= 4.

\subsection{\label{sec:FIQ:microlensing}Detecting exoplanets via gravitational microlensing}

\begin{figure}[b]
\includegraphics[width=\linewidth]{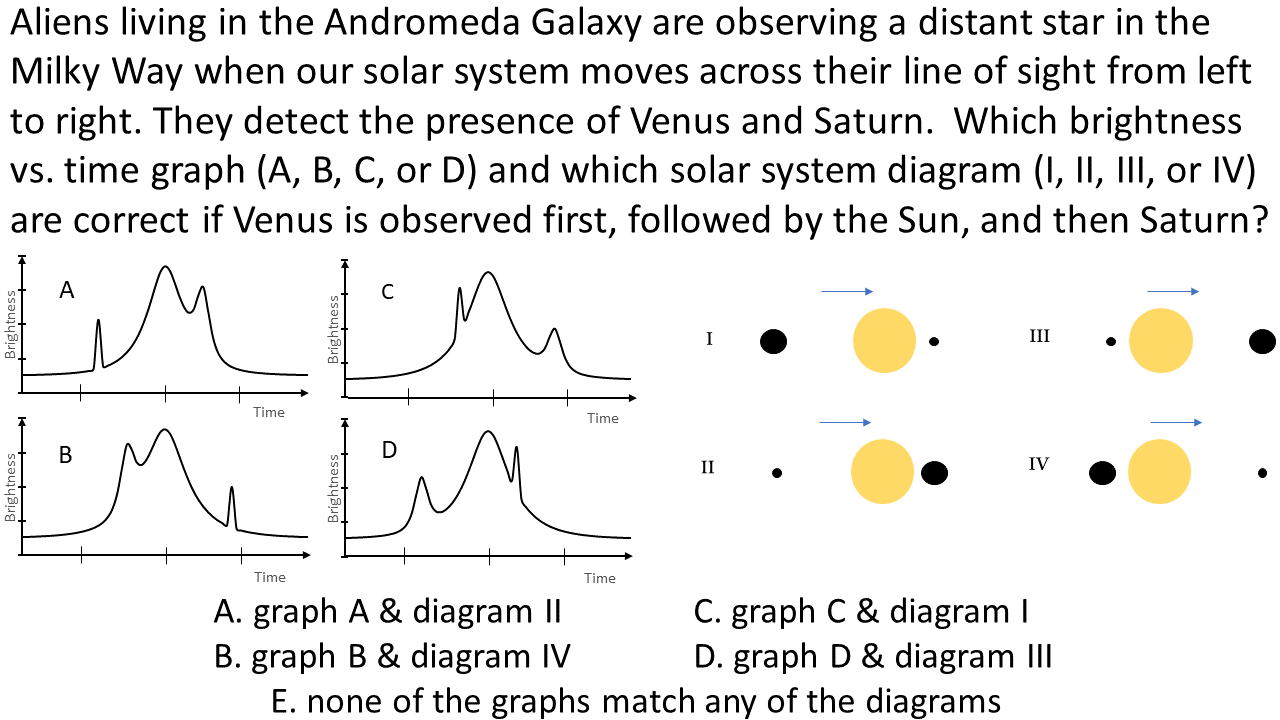}
\caption{\label{fig:FIQ:microlens}FIQ on detecting exoplanets via gravitational microlensing.}
\end{figure}

The Fluency-Inspiring Question in Fig.~\ref{fig:FIQ:microlens} 
uses a single-outcome matching format to focus a student's thinking. This question relies strongly on the information provided in the question stem, which clearly defines the single outcome and requires learners to both evaluate the graphs and diagrams, and find a matching set. The features and patterns of the graphs and diagrams, along with the combinations represented in the answer choices, are selected to emulate a wide range of possible outcomes and encapsulate the most common incorrect reasoning displayed by students during classroom testing on the topic \cite{WCPB2016}. As an example, students often believe that the left or right position of a planet relative to the companion star will be preserved in the left or right location (respectively) of the brightness peak caused by the planet on the brightness vs. time graph. These students do not always realize that, depending upon the system's direction of travel, a planet located to the right of a star may move across the field of view first, making the brightness peak for that planet occur first in time, which is to the left on the time axis.

What makes this question truly distinct as a Fluency-Inspiring Question is the reliance on specific Pedagogical Discipline Representations. For an expert in this field, the common discipline representations have their origins in general relativity and are dominated by advanced mathematics that would be completely opaque to the intended audience \cite{WCPB2016}. In this question (Fig.~\ref{fig:FIQ:microlens}), neither the graphs nor the diagrams are scientifically accurate at the level one would expect in a research paper, colloquium slide, or textbook treatment of the topic. Yet these graphs and diagrams have high pedagogical affordances -- they contain just the right mixes of highly contextualized bits of information to make these representations accessible to non-science majors trying to characterize the properties of a two exoplanet system discovered via gravitational microlensing. Since this question's scenario involves content that typically spans several weeks of a course, meaningful group discussions require students to unpack and connect a wide range of astrophysical topics, from solar system properties to general relativity, thus elevating this question to a QCR~= 4.

\section{\label{sec:SRTs}Student Representation Tasks}

Our \textit{curriculum development framework} helped us to create Fluency-Inspiring Questions, a more pedagogically powerful extension of a well-established active learning strategy. But perhaps a more noteworthy curricular object generated from the application of our curriculum development framework is a brand new type of active learning strategy, one not previously seen in astronomy instruction: \textbf{Student Representation Tasks} (SRTs). These fluency-inspiring active learning strategies are especially innovative in that they ``flip the script,'' making the learners responsible for creating the representations. This shifting of cognitive load, from interpreting to \textit{creating} representations, is a dramatic pedagogical design choice and was informed by the theoretical perspective of social semiotics. This unique shift in how representations are used to motivate fluency warrants specific examples that unpack how SRTs focus the learners' cognitive efforts.

In general, Student Representation Tasks are designed to intellectually engage learners by requiring them to evaluate and make connections between complex astrophysical relationships and reflect upon those ideas as they produce their own discipline representations to depict a specific physical scenario. The student-generated discipline representations commonly take the form of a diagram or sketch of the physical situation accompanied by labels, arrows, data tables, graphs, etc., that characterize the relevant astrophysics. \citet{TPAFG2019} suggest that drawing both promotes and influences reasoning in several ways: (1) reasoning occurs via the act of drawing, (2) the drawing itself facilitates further reasoning, and (3) the resulting drawing stands as a representation of reasoning. They also suggest that three conditions must be met in order for drawing to facilitate and support reasoning and scaffold learning  \cite{TPAFG2019}. First, the activity must be carefully structured such that the act of creating the representation(s) is viewed by the learners as a reasoning process, allowing them to interpret and explain one or more phenomena. The activity's structure, then, requires a sufficiently clear focus and carefully coordinated modes and associated affordances such that learners can successfully critique their own work while allowing for some diversity in the resulting drawings. Second, the students must already have the necessary background in both content knowledge and exposure to the relevant conventional representations. And finally, instructors must provide meaningful real-time and ongoing guidance, feedback, and scaffolding support.

From the perspectives of social semiotics, instructional design, and student engagement, the creation of Student Representation Tasks necessitates considering some of the requirements of our curriculum development framework (\S\ref{sec:framework-dev}) slightly differently. The first requirement guides development in essentially the same way since SRTs are still focused on one or more particular discipline topics and associated learning outcomes. SRT development is somewhat different, however, with regard to how framework requirements 2--5 play out. We still examine the canonical discipline representations, consider their pedagogical values, and carefully consider how other modes of representations and intellectual tasks might better align with the discipline topic's learning outcomes. However, in deciding on the modes of representation and creating the corresponding physical situation that will meaningfully connect them, we must consider that for an SRT, \textit{ultimately it is the students who will create the representations}. The activity design must then orchestrate students working in groups to engage in the discourse that (hopefully) leads to the thoughtful creation of appropriate discipline representations. In this way, Student Representation Tasks are quite different from most active learning activities in astronomy as \textit{they explicitly shift the role of generating representations onto the students} -- a shift that fosters unique opportunities from both a pedagogical design and a student learning standpoint.

Student Representation Tasks are post-instruction activities (in our case this means post-lecture and after implementing other active learning strategies such as Lecture-Tutorials, Ranking Tasks, and Think-Pair-Share) designed to be completed in small collaborative groups of two or three students. To begin, students are given a problem statement that provides essential details about a highly contextualized physical scenario along with key information regarding the desired representations. The peer-to-peer discussion fostered by an SRT brings about a negotiation of choices the students must make in order to create the representations. This disciplinary discourse is the vehicle that furthers the development of fluency with the topic.

Next, we highlight two Student Representation Tasks to provide insight into how our curriculum development framework leads to particular instructional design choices and pedagogical outcomes. The first example, on the topic of Doppler shift, shows how an SRT's instructions and context can motivate the creation of several different representations (all on the same topic) to result in the construction of a more complex, final representation that confronts known student conceptual and reasoning difficulties. In the second example, we provide students with an opportunity to think about one of the most intriguing ideas in science, \textit{lookback time}, and the notion that telescopes serve as time machines, allowing us to observe events that occurred in the past. Additionally, this activity gives students the chance to connect lookback time with their knowledge about stellar properties, stellar evolution, and a canonical discipline representation that is used extensively in nearly every astronomy course: the Hertzsprung-Russell diagram.

\subsection{\label{sec:SRT:doppler}Doppler shift}

\begin{figure}[htbp]
\includegraphics[width=\linewidth]{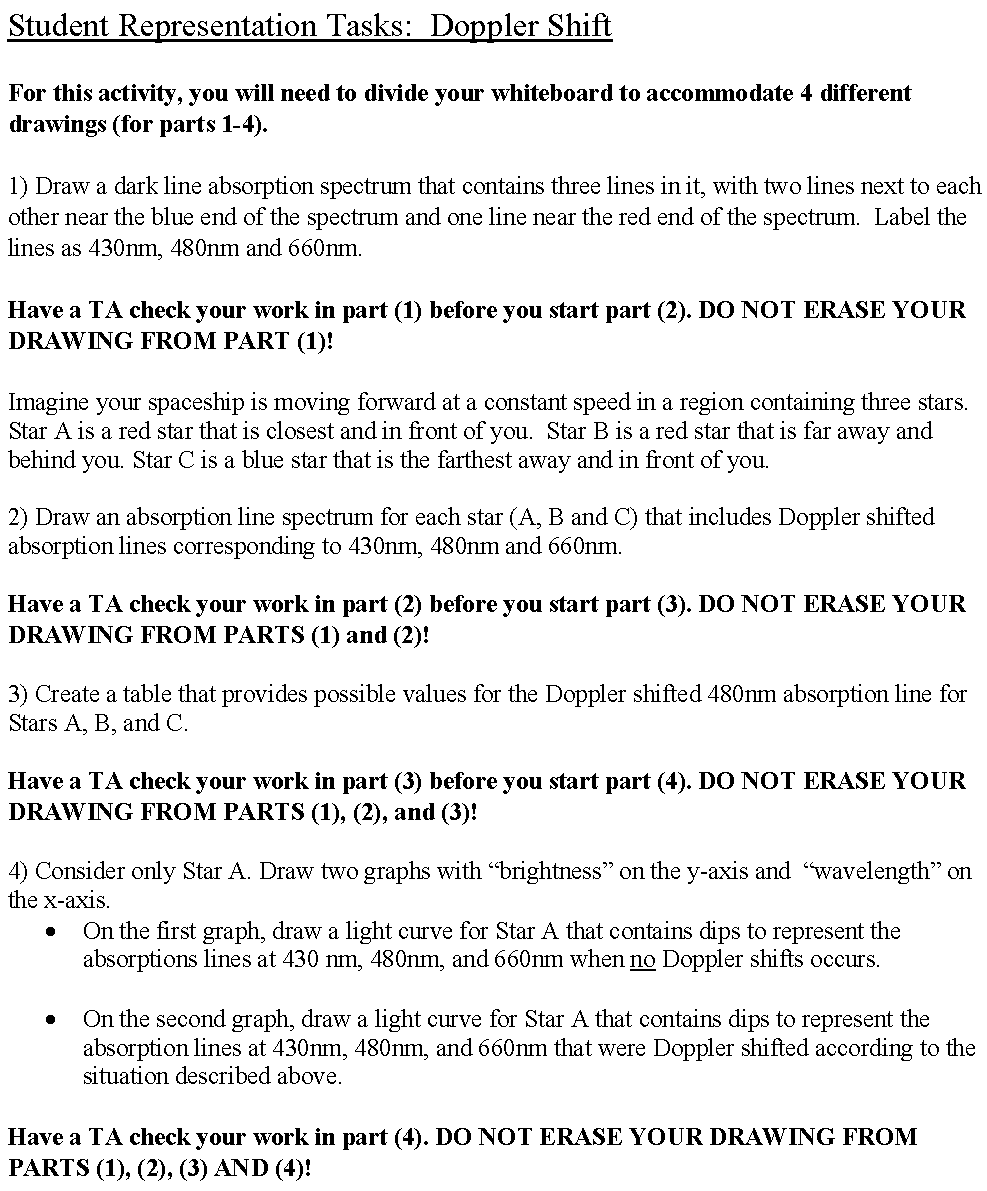}
\caption{\label{fig:SRT:doppler}SRT on Doppler shift.}
\end{figure}

Students' prior knowledge about, and experiences with, Doppler shift can make it surprisingly challenging for them to comprehend that the phenomenon is due only to the relative motion between the source and observer. Even after instruction on the topic, this struggle persists. Our Doppler shift Student Representation Task (Fig.~\ref{fig:SRT:doppler}) 
targets some of the key misconceptions and reasoning difficulties encountered when helping students become fluent with the concept.

The sequence of prompts in the Doppler shift SRT asks students to first create sketches of dark line absorption spectra, then make a data table, and finally sketch energy output vs. wavelength graphs. Students are given information that is physically relevant and information that targets key na\"{i}ve ideas and reasoning difficulties, such as relating Doppler shift to a star's color or distance. It is important to note that neither the students' na\"{i}ve ideas nor the usual course treatment of the topic (at this point in a typical term's sequence) are connected to photometric or cosmological redshifts, the Hubble-Lema\^{i}tre Law, or interstellar reddening.

While the first representation provides sufficient challenge to those students with only a rudimentary understanding, it is in doing the later parts of this activity that students begin to engage with their peers at deeper levels. Note that in the second part of the activity we provide information that is explicitly chosen to confront the two most common difficulties students have when reasoning about physical situations involving the Doppler shift.

Our decades of classroom experience teaching Doppler shift shows us that many students invoke an incorrect relationship between an object's color and the color label used in science to identify the change in wavelength due to the relative motion between the source and observer. These learners incorrectly reason, for example, that a star that is actually red must be moving away from the observer, or that a star whose light is redshifted must actually be red in its true color. Additionally, some students incorrectly associate an object's distance with the directional sense of the Doppler shift and with the color of the wavelength change, reasoning that a star moving toward the observer must be closer than a star that is moving away, or that the shift to a shorter, ``bluer'' wavelength implies that the object must be closer than one whose light is shifted to longer, ``redder'' wavelengths. The information provided in the activity regarding a star's color, distance, and position relative to the moving observer are deliberately varied in order to provide student groups with sufficient contrasting cases that address the different Doppler-specific reasoning difficulties. Although not explicitly called for, we find that a majority of student groups choose to make drawings of the physical scenario described in the second part -- a sign that they find the representation valuable in, and possibly necessary to, facilitating their unpacking and discernment of this phenomenon. 

In the third part of the Doppler shift SRT, students must estimate a numerical wavelength for one of the shifted absorption lines associated with each of the three stars while considering the other information about color, distance, and relative position (direction of motion). We find this to be a highly discriminating task, and one that students rarely, if ever, encounter in traditional instruction or active learning activities on this topic (including those we have previously developed).

The final part of this Student Representation Task requires students to create two energy output vs. wavelength graphs for one of the three stars. This is an important intellectual engagement opportunity for students who incorrectly reason that only the absorption line at the end of the spectrum corresponding to the color of the light associated with the shift (redshift or blueshift) will be shifted from its ``rest'' wavelength. Students are also challenged to correctly draw what is essentially a blackbody curve such that the peak is correct for the actual color of the star (Wien's law) and the three dips (absorption features) are repositioned relatively correctly in accordance with the sense of Doppler shift occurring (``red'' or ``blue''). It is this part of the SRT that elevates the activity to a QCR~= 4.

\subsection{\label{sec:SRT:lookback}Lookback times and stellar properties}

\begin{figure}[b]
\includegraphics[width=\linewidth]{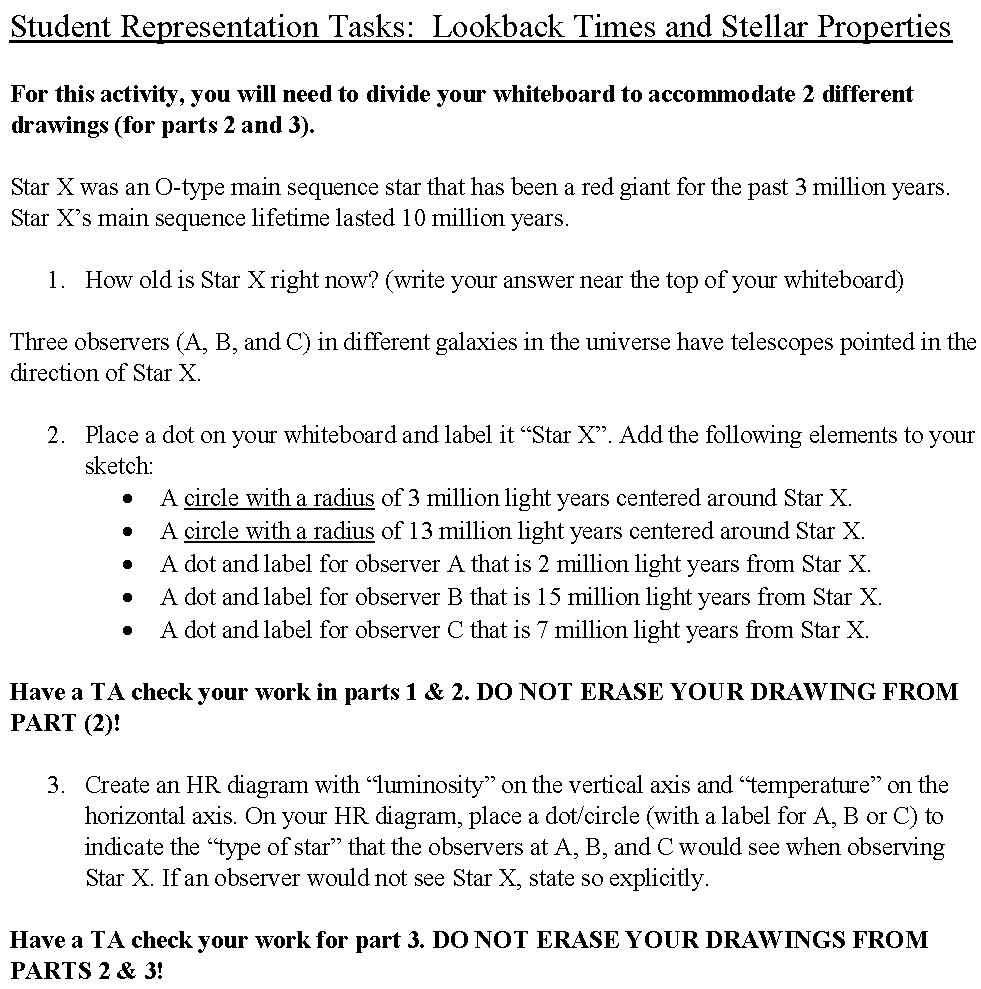}
\caption{\label{fig:SRT:lookback}SRT on lookback times and stellar properties.}
\end{figure}

The information provided in the Lookback Times and Stellar Properties Student Representation Task (Fig.~\ref{fig:SRT:lookback}) 
includes the evolutionary state of a star, its spectral type, its main sequence lifetime, and distances from the star to three observers in the universe. The provided information establishes a particular set of outcomes for different locations in the universe. We first ask students to draw two circles centered on the star, one with a radius of 3 million light-years and another of 13 million light-years -- distances that correspond to the light travel time to the beginning of the star's main sequence phase and to the beginning of its red giant phase. Observer distances are chosen such that one observer sees the star as a (high mass) main sequence star, another sees it as a red giant, and another sees nothing. This combination of parameters creates a set of potential circumstances that allows us to differentiate between students who have a robust model of lookback time and those who believe that the distance between the star and an observer relates to the time since the beginning of the star's life rather than the time back from the star's current state of existence. By making students reason simultaneously about lookback time, stellar evolution, and stellar properties on a Hertzsprung-Russell (H-R) diagram, this SRT provides students the opportunity to engage at the QCR~= 4 level.

Instructors are quickly able to discern the reasoning pathways students use by seeing how they depict on their H-R diagrams the different evolutionary states each observer will see for Star X. That is, a student who places a dot (or large blue circle) to indicate a main sequence star that is both luminous and hot for the observer at location A, and a dot (or large red circle) to indicate a star that is both luminous and cool for the observer at location C, is simultaneously indicating the correct evolutionary states while demonstrating an incorrect understanding of how lookback time changes how (and when) the observer perceives an object or event.

\section{\label{sec:concl}Conclusions}

We highlight the theoretical perspectives that inform our past curriculum development work and the development of our \textit{curriculum characterization framework} used to identify the variety of modes of representation and intellectual tasks used in, as well as the levels of disciplinary discourse required to explain the reasoning behind one's answers to, an active learning activity. We briefly discuss the application of our curriculum characterization framework to systematically code 353 faculty-produced multiple-choice Think-Pair-Share questions and the insights this work provides into the decisions faculty make when given the opportunity to design curriculum in a professional development setting. This investigation revealed that, for some astronomy and physics topics, there appears to be an over-reliance on a small assortment of discipline representations and intellectual tasks, and a predisposition towards relatively low levels of complexity. We acknowledge that our results may be incomplete as the professional development workshops' settings had several limitations that may have restricted faculty members' abilities to generate higher level questions. Additionally, the faculty who produced these questions are not representative of all faculty, as many were in the early stages of their careers, some with little to no teaching experience.

What we learned from our investigation into these Think-Pair-Share questions informed the creation of a second framework -- our \textit{curriculum development framework} -- that is useful for generating active learning strategies that move students towards discipline fluency by creating rich opportunities for students to practice discernment, unpacking, reflection, and metacognition.

We used our curriculum development framework to design \textbf{Fluency-Inspiring Questions}, which help students make robust connections amongst a complex set of complementary discipline representations, cognitive tasks, and discipline-specific ideas. We also provide insight into another new type of active learning activity generated using this framework -- \textbf{Student Representation Tasks}. SRTs shift the responsibility of creating discipline representations onto the shoulders of the learners and are pedagogically very powerful, both in terms of the richness of the student learning experience fostered and the discriminatory abilities these tasks offer the instructor with regard to revealing what has or has not been learned through prior instruction.

Developing and using our two frameworks also informs how we, as instructors and researchers, think about our disciplines, and provides a pathway for exploring more pedagogically interesting and powerful opportunities to help learners develop their discipline fluency. Our iterative and continuously evolving process of research-informed curriculum development -- supported by numerous theoretical perspectives -- leads to innovations in the development and assessment of active learning strategies and provides new insights into the teaching and learning of physics and astronomy. It is exciting to note that sharing preliminary versions of our frameworks with the DBER\footnote{\underline{D}iscipline \underline{B}ased \underline{E}ducation \underline{R}esearch} community has (1) led to collaborations with faculty who were inspired to create their own FIQs and develop new curricular materials on topics as yet unaddressed with active learning strategies, and (2) opened up dialogues with faculty outside of physics and astronomy about the extension and application of our frameworks to their disciplines.

\begin{acknowledgments}
Many thanks go to Nate Goss for his substantial input into numerous sessions iterating on and refining the coding schemata, and to Gina Brissenden for her assistance compiling the questions from the various workshops. We are also grateful to three referees for their thoughtful comments that vastly improved this manuscript.
\end{acknowledgments}

\providecommand{\noopsort}[1]{}\providecommand{\singleletter}[1]{#1}%

\end{document}